# Fast Spiking of a Mott $VO_2$-Carbon Nanotube Composite Device

*Stephanie M. Bohaichuk[1#], Suhas Kumar[2*#], Greg Pitner[1], Connor J. McClellan[1], Jaewoo Jeong[3], Mahesh G. Samant[3], H-.S. Philip Wong[1], Stuart S. P. Parkin[3], R. Stanley Williams[4], Eric Pop[1,5]*

[1]Stanford University, Electrical Engineering, Stanford, CA 94305, USA
[2]Hewlett Packard Labs, 1501 Page Mill Rd, Palo Alto, CA 94304, USA
[3]IBM Almaden Research Center, 650 Harry Road, San Jose, CA 95120, USA
[4]Texas A&M University, Electrical & Computer Engineering, College Station, TX 77843, USA
[5]Stanford University, Material Science & Engineering, Stanford, CA 94305, USA

*Address correspondence to S.K. (su1@alumni.stanford.edu). [#]Equal contribution

*The recent surge of interest in brain-inspired computing and power-efficient electronics has dramatically bolstered development of computation and communication using neuron-like spiking signals. Devices that can produce rapid and energy-efficient spiking could significantly advance these applications. Here we demonstrate DC-current or voltage-driven periodic spiking with sub-20 ns pulse widths from a single device composed of a thin $VO_2$ film with a metallic carbon nanotube as a nanoscale heater. Compared with $VO_2$-only devices, adding the nanotube heater dramatically decreases the transient duration and pulse energy, and increases the spiking frequency, by up to three orders of magnitude. This is caused by heating and cooling of the $VO_2$ across its insulator-metal transition being localized to a nanoscale conduction channel in an otherwise bulk medium. This result provides an important component of energy-efficient neuromorphic computing systems, and a lithography-free technique for power-scaling of electronic devices that operate via bulk mechanisms.*

The emergence of artificial intelligence and data-intensive tasks has necessitated a revamp of computing hardware beyond transistor-based Boolean logic and the von Neumann architecture.[1,2] Within this revamping effort lies the broad domain of neuromorphic computing which aims to exploit biologically-inspired processes, namely computing, communicating, and operating a neural network using electrical spiking.[3-8] In order to improve the energy-efficiency and speed of such systems it is desirable to control the pulse width and energy, and to produce the spiking using single scalable devices.[4,9-11] For instance, adjusting the analog node weights of a neural network by small increments in order to enable high precision will require precise and tunable low energy pulses, especially in networks that use memristors such as phase change memory or oxide ionic resistive switches.[12,13]

Partly owing to the absence of compact circuits that can produce such tunable low-energy pulses, even the best memristor-based neural networks have had to implement elaborate transistor-based circuits at every node of very large networks, making the system's efficiency far from ideal.[14] Instead, compact spiking systems without transistors can be constructed by exploiting transient dynamics and/or electronic instabilities, for instance, the temporally abrupt resistance changes during a Mott insulator-metal transition (IMT) causing a capacitive discharge.[15] $VO_2$ and $NbO_2$ are widely studied Mott insulators that undergo IMT above room temperature (~340 K in $VO_2$, ~1100 K in $NbO_2$), resulting in abrupt changes of their electrical resistance (usually by several orders of magnitude) often accompanied by measurable electrical instabilities such as negative differential resistance (NDR) as increasing current is applied.[16,17] Relaxation oscillators that are constructed with $NbO_2$ or $VO_2$ typically use a DC-current or voltage driven Pearson-Anson-like circuit that includes a capacitor in parallel with the NDR element, and a suitable series resistor.[15,18-22] However, generation of fast DC-voltage-driven periodic spiking will require aggressive scaling to minimize both the electrical and thermal time constants of these oscillators,[11,23] for instance, by careful electron-beam lithography, which may not be practical for large-scale electronics.

Here we utilize a metallic single-wall carbon nanotube (CNT) of ~1 nm diameter as a nanoscale heater in contact with a thin film of $VO_2$, in order to effectively scale a micrometer-sized device to sub-10 nm width.[24] A single CNT-$VO_2$ composite device with an integrated parallel capacitance from the contact electrodes forms a relaxation oscillator when driven by a DC source, producing periodic spiking. The effective scaling of the device width using a CNT yields dramatic improvements in the dynamic spiking behavior, including an increase in frequency and a reduction of pulse energy and transient time scales by nearly three orders of magnitude compared to $VO_2$ control devices without a CNT. We further demonstrate that the frequency, duty cycle, and pulse width could be tuned within a single CNT-$VO_2$ device by nearly one order of magnitude by altering the DC bias conditions.

The device structure (Figures 1a-1c) contains a lateral active region defined by a ~5 nm thin strip of $VO_2$ contacted by Pd electrodes, essentially forming a planar IMT device. To construct the CNT-$VO_2$ composite device, aligned CNTs were grown on a quartz substrate then transferred onto the surface of $VO_2$ before the device structure was defined (Figures 1d-1f) (see Supplementary Material for details of device fabrication and material properties). The quasi-static current-voltage behavior was measured using a current sweep (Figures 1g-1h) for both $VO_2$ and CNT-$VO_2$ devices. With an otherwise identical geometry ($L$ = 5 μm, $W$ = 4 μm), both exhibited qualitatively similar regions of NDR along with hysteresis. These behaviors are characteristic of a thermally-activated transport mechanism and a temperature-controlled Mott transition, detailed elsewhere.[16,25] The quantitative differences between the two devices arise from a much narrower "thermal width" in the CNT-$VO_2$ device, which also increased the range of currents for which NDR occurs. The CNT acts as an ultra-narrow nanoscale heater,[26] reducing the heated volume (thermal capacitance) of the $VO_2$. We use three-dimensional (3D) finite element electrothermal simulations (Figure 2) to illustrate that just prior to switching, the volume of $VO_2$ heated close to its transition temperature reduces by three orders of magnitude upon introduction of a CNT. The simulations also show that the effective thermal capacitance reduced by 3 orders of magnitude. The thermal capacitance affects the dynamical behavior significantly more than the quasi-static characteristics, similar to how an electrical capacitance affects the dynamics more than the static behavior. The threshold quasi-static power to trigger NDR and switching was reduced by a factor of 2-5× in a CNT-$VO_2$ compared to a $VO_2$-only device[24] while, as we will demonstrate, the dynamical behavior is affected by several orders of magnitude.

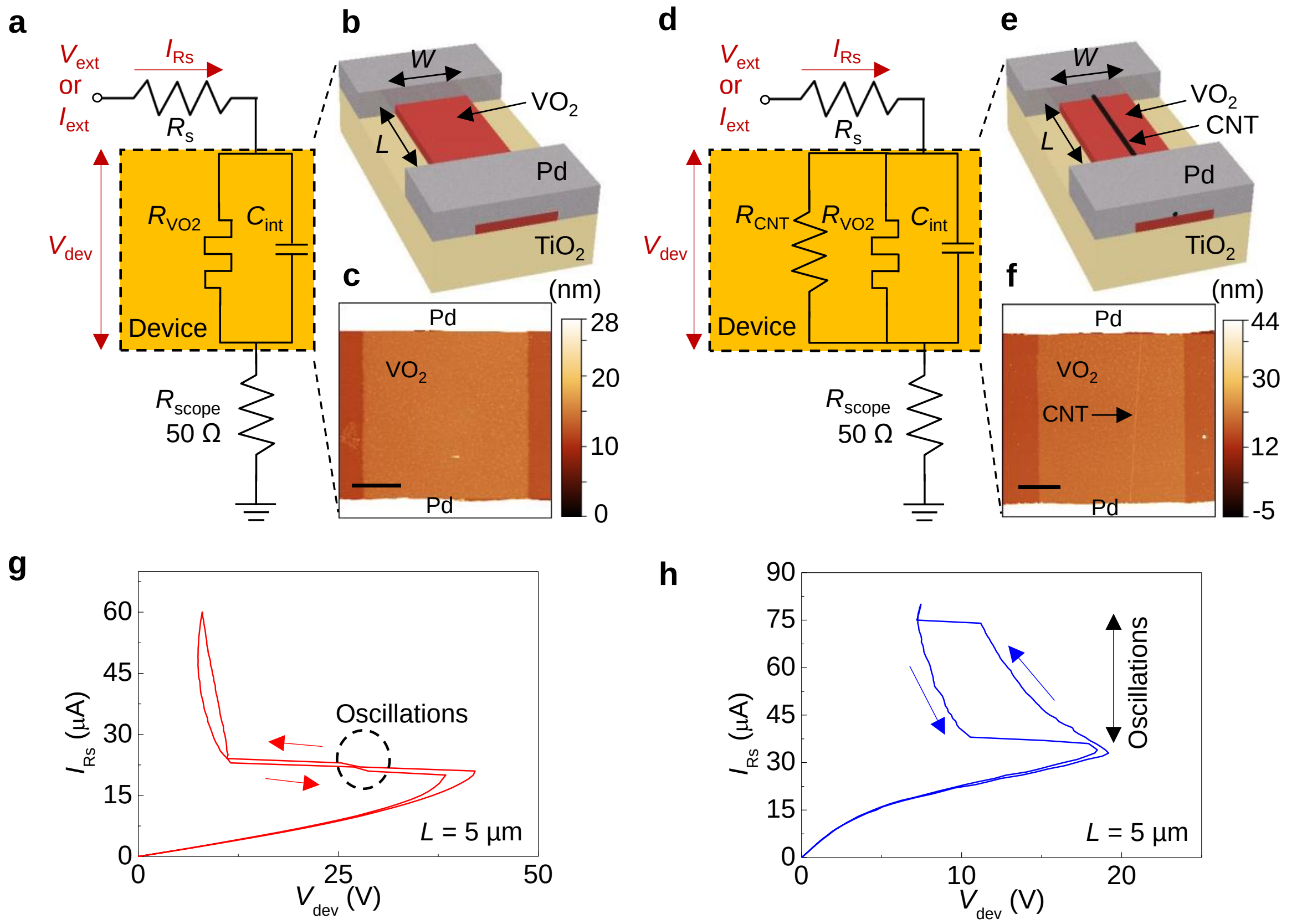

**Figure 1: Device structure and static behavior.** (a-c) $VO_2$-only control device, schematic and atomic force microscopy (AFM) image of fabricated device (scale bar is 1 µm). The circuit used for all static and dynamic measurements is also shown; the static measurements included the oscilloscope. (d-f) CNT-$VO_2$ composite device, schematic, and AFM image. (g) Measured quasi-static current-voltage behavior of a $VO_2$-only device, using a current source. (h) Similarly measured behavior of a CNT-$VO_2$ device. Biasing at a constant current in the NDR region results in oscillations.

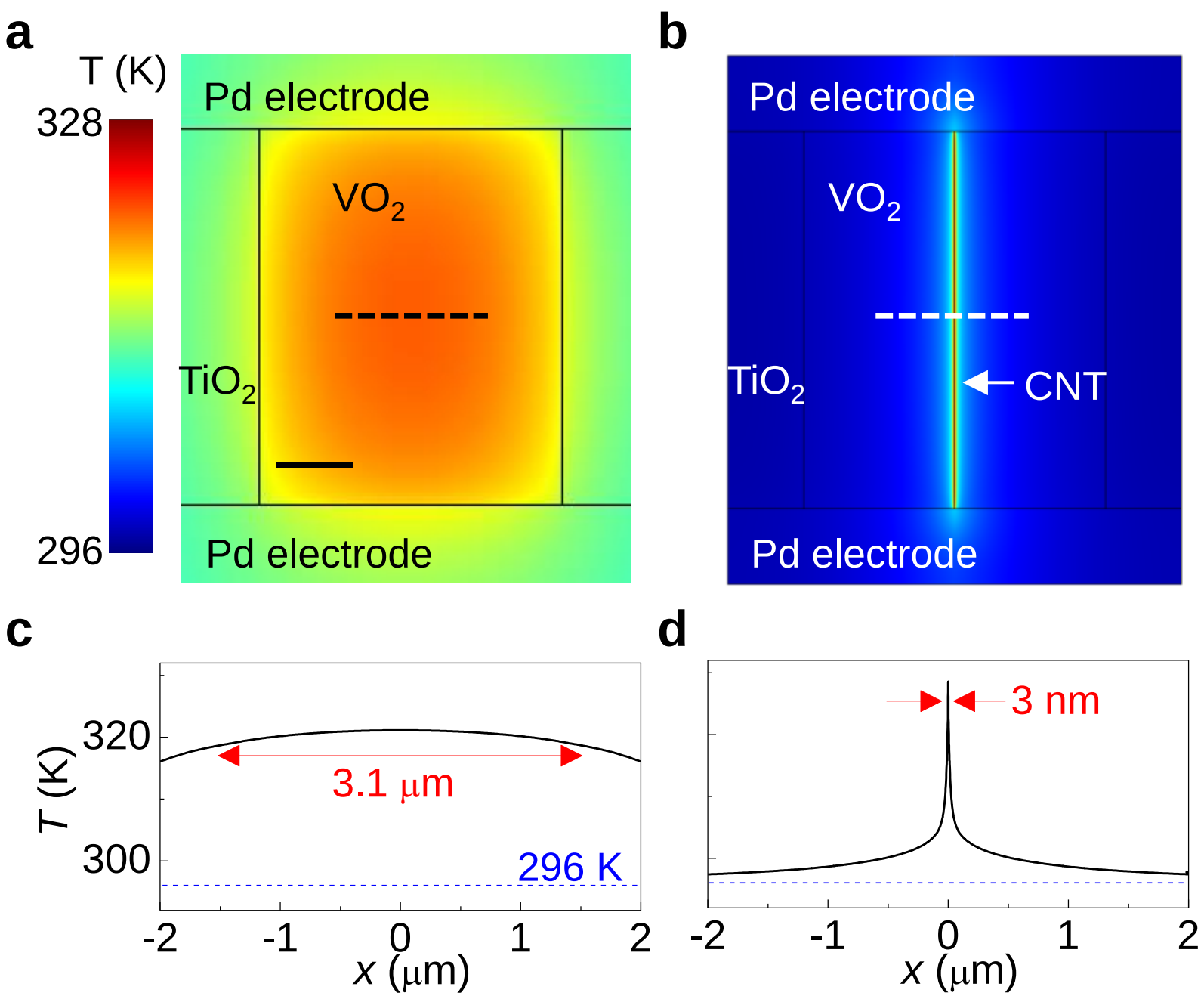

**Figure 2: Simulated static temperature maps.** Static surface temperature maps obtained by three-dimensional electrothermal finite element simulations of (a) a $VO_2$-only device, and (b) a CNT-$VO_2$ device at a DC voltage bias just below the switching threshold (both with $L$ = 5 μm, $W$ = 4 μm). Scale bar is 1 μm. (c)-(d) Temperature profiles along the dashed horizontal lines in (a) and (b), respectively. Blue dashed horizontal lines indicate room temperature. Red arrows indicate the width heated to 90% of the maximum temperature increase above the ambient, showing a reduction by three orders of magnitude in the CNT-$VO_2$ device, compared to a $VO_2$-only device. This measure is used to illustrate the significant reduction in the thermal mass upon introducing a CNT.

To study the dynamical behavior of the two circuit elements, we applied a constant external current to each device within its NDR region. Because of the intrinsic capacitance of the devices and the large series resistance of the current source, a Pearson-Anson-like relaxation oscillator was formed. This produced a periodic train of repeating pulses or spikes, typical of Mott memristors, due to the periodic heating and cooling of $VO_2$ across its IMT, which controls the periodic discharging and charging of the capacitor.[27] Figures 3a-3b display the periodic spiking of a $VO_2$-only and a CNT-$VO_2$ device that were nominally identical in geometry ($L$ = 5 μm, $W$ = 4 μm) except for the inclusion of the CNT in the latter. In the $VO_2$-only device (Figures 3a, 3c) each spike consisted of an initial spike with a full width at half maximum (FWHM) of 28.5 ns (inset of Figure 3c), followed by a longer transient lasting 0.68 ms. Fitting an exponential decay to the falling edges yielded a time constant of 17 ns for the initial spike, and 0.24 ms for the longer transient (analyses detailed in Supplementary Material Figure S8). In contrast, the CNT-$VO_2$ device (Figures 3b, 3d) showed a spike with a FWHM of 36.5 ns and single decay constant of 27 ns, with no apparent longer transient. Thus, although the CNT-$VO_2$ device exhibited a spike comparable in duration to the initial large spike of the $VO_2$-only device, the former exhibited nearly negligible transient dynamics following the spike. This resulted in a 1000-fold increase in frequency (~0.5 MHz in CNT-$VO_2$ and 0.3 kHz in a $VO_2$-only device), and a 100-fold reduction in total pulse energy (1.3 pJ in CNT-$VO_2$ and 100 pJ in a $VO_2$-only device). We used an indirect experimental technique (via polymer-coating of the devices, detailed in Supplementary Material Section 5) to qualitatively confirm that (a) the power dissipation and transient widths of the metallic phase of the $VO_2$ were much smaller upon introduction of a CNT, (b) there were two time constants, a larger-amplitude faster spike and a smaller-amplitude slower transient, and (c) heating indeed occurred in the same location as the CNT.

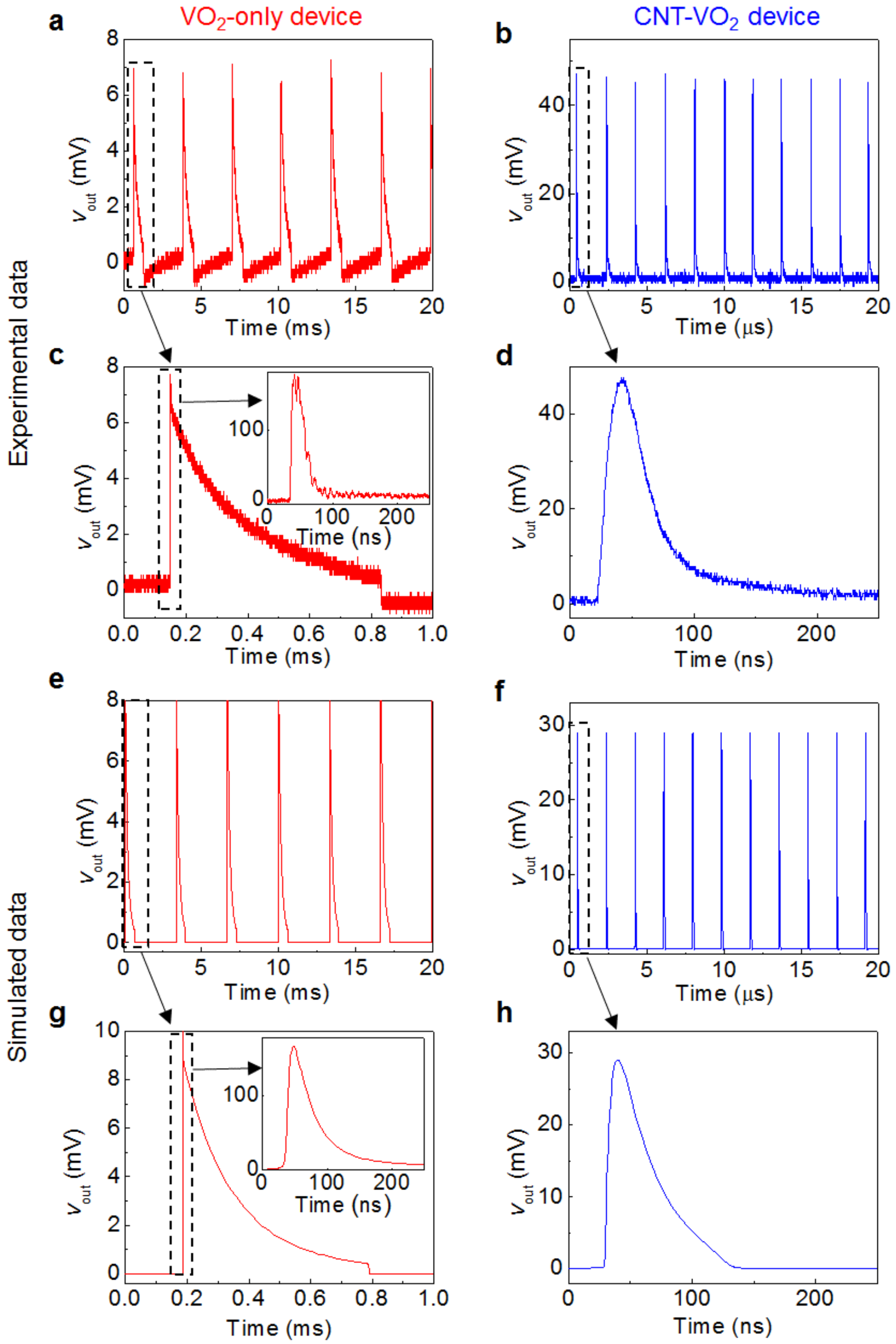

**Figure 3: Experimental and simulated periodic spiking**. Measured periodic voltage spiking of (a) a $VO_2$-only device driven by a 23 μA DC current and (b) a CNT-$VO_2$ device driven by a 60 μA DC current. (c) Magnified plot of a single pulse in the $VO_2$-only device, as indicated with a dashed rectangle in (a). The inset is a magnified plot of the transient spike at the rising edge of the pulse. (d) Magnified plot of a single pulse in the CNT-$VO_2$ device, as marked with a dashed rectangle in (b), which is missing the long transient tail evident in (c). (e)-(h) are simulation results, in comparison with (a)-(d). All abscissa (time) values were arbitrarily offset for clarity of presentation. All insets have the same ordinate units as the corresponding panel.

In order to gain insight into the effects of using a CNT to scale an electronic device down, we constructed a simplified device model that was incorporated into a relaxation oscillator circuit. The device model consisted of thermally-activated Schottky transport along with Newton's law of cooling representing temperature dynamics, which together are known to produce instabilities such as NDR (detailed in the Supplementary Material Section 2).[28,29] We used this model to reproduce the shape, frequency and the durations of the dynamics (Figures 3e-3h) by altering only one parameter upon introduction of the CNT: the lumped thermal capacitance was reduced by about three orders of magnitude to account for the reduced volume of $VO_2$ being heated across its IMT (consistent with the finite element simulations). The agreement of our model with experimental data is remarkable, given the simplicity of the model, and confirming our hypothesis that the CNT is indeed dramatically scaling down the effective thermal mass, the full effect of which can be inferred only by measuring the dynamics.

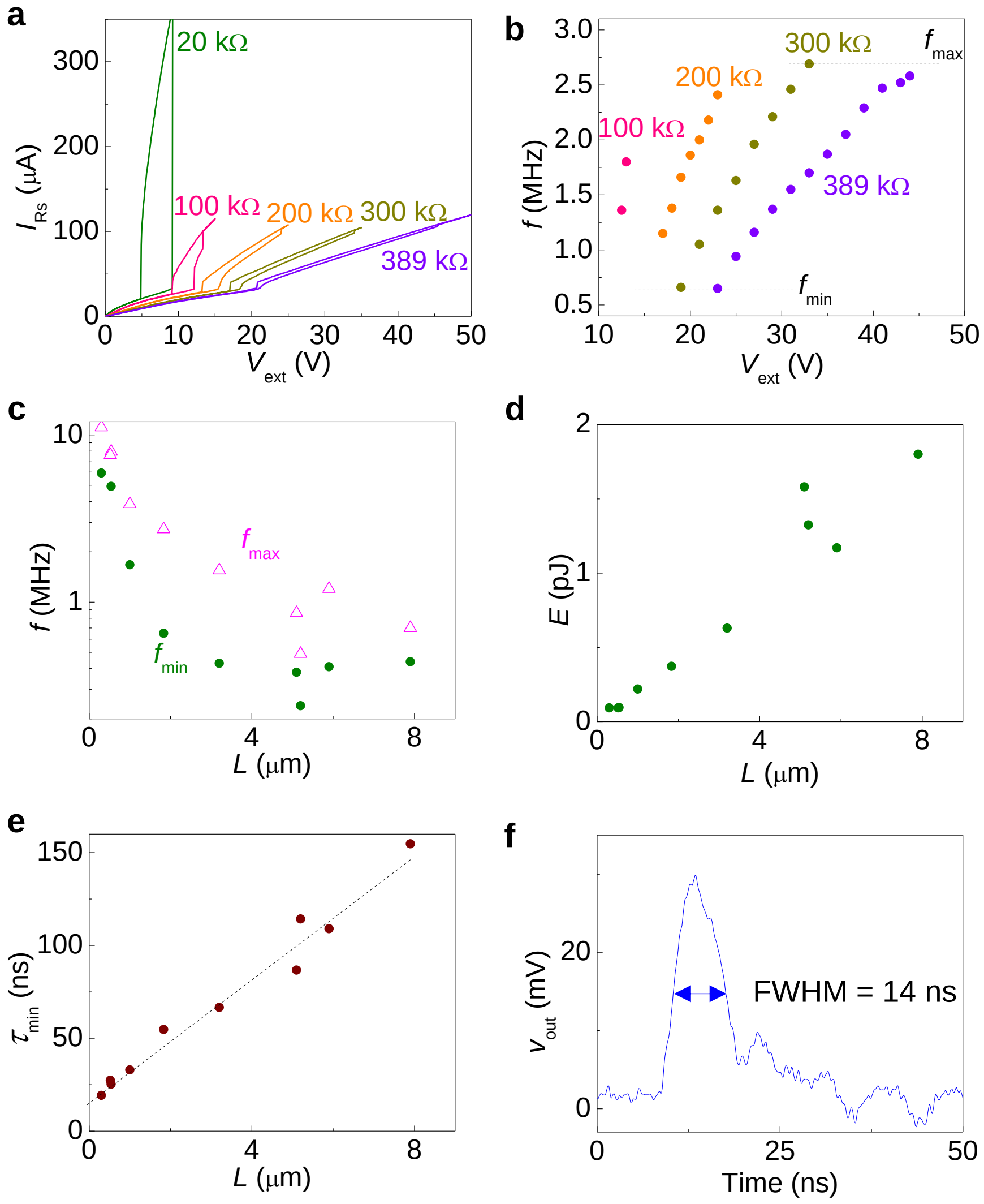

**Figure 4: Periodic spiking behavior varying with device and circuit parameters.** (a) Quasi-static current-voltage behavior of a single CNT-$VO_2$ device with dimensions $L$ = 1.8 μm and $W$ = 6 μm, measured using a voltage source and different series resistances, as marked in the color-coded legend. (b) Oscillation frequencies resulting from biasing the device at specific voltages on the quasi-static behavior characteristics in (a). The highest frequencies occur at the highest voltages, but slightly shorter pulse widths occur at lower voltages and frequencies. (c)-(e) Dependence of maximum and minimum frequency, minimum energy delivered by a single pulse, and the minimum width (defined as the full width between 10% of the maximum) of a single pulse corresponding to varying lengths of different CNT-$VO_2$ devices. (f) Plot of a single pulse within periodic spiking corresponding to the shortest device ($L$ = 0.3 μm), at the minimum of the trends in (d) and (e).

Within a single CNT-$VO_2$ device, we studied the range of applied DC voltages and corresponding series resistances for which the device exhibited periodic spiking. As is expected from the theory of local activity and nonlinear dynamics,[27] we observed that the range of voltages for which periodic spiking occurred increased with increasing series resistance (Figures 4a-4b), with the frequency range spanning nearly an order of magnitude. We further characterized the periodic spiking behavior across CNT-$VO_2$ devices with varying lengths between the electrodes. The maximum and minimum frequencies, minimum pulse width, and the minimum energy delivered by a pulse, all spanned an order of magnitude as the device length was scaled by an order of magnitude (from 300 nm to 8 μm), displayed in Figures 4c-4e. In particular, the shortest device ($L$ = 300 nm) exhibited the shortest pulse width, with a FWHM of ~14 ns, and a total energy delivered by the pulse (to a 50 Ω oscilloscope load) of ~93 fJ (Figure 4f). The shortest pulse width corresponded to the lowest of the measured frequencies within the device (~6 MHz). The shortest device also exhibited a maximum frequency of 11 MHz, the highest of all devices measured. The linear trend in the minimum pulse width (Figure 4e) had a non-zero $y$-intercept of ~15 ns, which likely represents the minimum possible pulse width within the present measurement setup and device structure (a combination of parasitic time constants and contact resistances). Additional analysis on the periodic spiking and the individual pulses (rise and fall times) are provided in the Supplementary Material Section 4.

In conclusion, these results present a pathway to fast and low-energy neuron-like spiking with high demand in several types of neuromorphic computing applications. We have demonstrated that using a CNT nanoscale heater we can locally confine the effective thermal structure of an otherwise bulk behavior, thereby engineering the dynamical properties by several orders of magnitude without needing extensive and time-consuming nanoscale lithography. This provides a promising platform to scale and control the behavior of many electronic components which rely on temperature, given the increasing use of temperature-driven functions such as NDR, IMT, and chaos. Apart from the scaling of periodic spiking studied here, several other applications benefit from such thermal engineering, for instance within memory selectors which minimize the standby power consumption of data storage, while utilizing a thermal runaway to trigger threshold switching.[30]

**Acknowledgements:**

The authors gratefully acknowledge Kye Okabe, Gary A. Gibson, Saurabh Suryavanshi and Aditya Sood for helping with the experimental setup, simulation codes, calculations and/or commenting on the manuscript. This work was supported in part by the Stanford SystemX Alliance and by the National Science Foundation (NSF). The research is also based upon work supported by the Office of the Director of National Intelligence (ODNI), Intelligence Advanced Research Projects Activity (IARPA), via contract number 2017-17013000002. Work was performed in part at the Stanford Nanofabrication Facility and the Stanford Nano Shared Facilities which receive funding from the National Science Foundation as part of the National Nanotechnology Coordinated Infrastructure Award ECCS-1542152. S.B. acknowledges support from the Stanford Graduate Fellowship (SGF) program and the NSERC Postgraduate Scholarship program.

# Fast Spiking of a Mott $VO_2$-Carbon Nanotube Composite Device

*Stephanie M. Bohaichuk[1#], Suhas Kumar[2*#], Greg Pitner[1], Connor J. McClellan[1], Jaewoo Jeong[3], Mahesh G. Samant[3], H.-S. Philip Wong[1], Stuart S. P. Parkin[3], R. Stanley Williams[4], Eric Pop[1,5]*

[1]Stanford University, Electrical Engineering, Stanford, CA 94305, USA
[2]Hewlett Packard Labs, 1501 Page Mill Rd, Palo Alto, CA 94304, USA
[3]IBM Almaden Research Center, 650 Harry Road, San Jose, CA 95120, USA
[4]Texas A&M University, Electrical & Computer Engineering, College Station, TX 77843, USA
[5]Stanford University, Material Science & Engineering, Stanford, CA 94305, USA

## Supplementary Material:

**Contents:**

### 1. Operation of the Relaxation Oscillator

A relaxation oscillator (such as a Pearson-Anson oscillator) exploits the multi-stabilities within memristive devices such as negative differential resistance (NDR) to generate self-oscillations upon biasing with a constant voltage. Similar to how a 555 timer integrated circuit emulates multi-stabilities using elaborate transistor circuits, and can therefore be used to produce oscillations upon biasing with a constant voltage, NDR devices exhibit such multi-stabilities within a single compact device. A typical working of a relaxation oscillator constructed using an NDR element ($M$) is depicted in Figure S1. A series resistor $R_s$ is included to fix an operating point on $M$ within the NDR region (green line in Figure S1a). Upon application of a DC input voltage, the capacitor in the circuit ($C_p$) gets gradually charged, and the voltage across the parallel NDR memristive element ($v_{Cp} = v_M$) also gradually increases, along with a gradual increase in the current through $R_s$. When $v_M$ reaches the threshold voltage ($V_2$), the NDR element immediately allows flow of a large current through itself, which causes a sudden jump in current through $R_s$. The creation of a conduction path within $M$ causes $C_p$ to gradually discharge and therefore gradually reduce $v_M$. This causes an exponential decay of current through $R_s$ in the 'ON' duration. When $v_M$ drops below the second threshold ($V_1$), $M$ immediately prevents the flow of a large current through itself. The cycle then begins again and repeats itself indefinitely, thereby creating oscillating signals. As an extension, since a current source is essentially a voltage source with a practically infinite series resistance, biasing with a current source within the region of NDR will also produce oscillations. It is notable that any physical process that can lead to an instability can be utilized to construct a relaxation oscillator, for instance, a Mott transition, super-linear thermal feedback, tunneling, etc. [*Nanotechnology 23, 215202, (2012); Applied Physics Letters 108, 023505, (2016); Advanced Functional Materials, 28, 1704175 (2018)*]

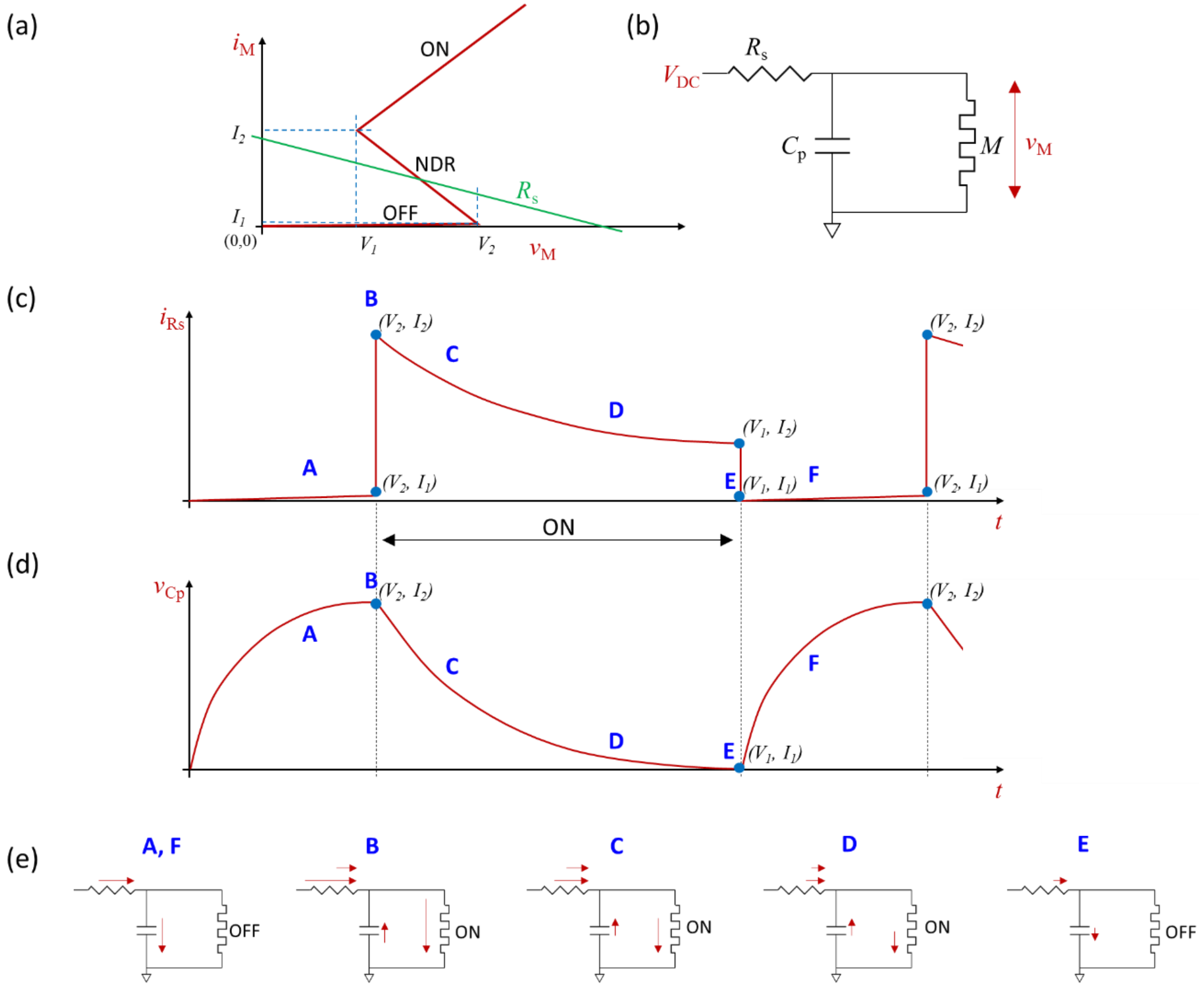

**Figure S1:** Working of a relaxation oscillator. (a) Quasi-static current-voltage behavior of a memristive device exhibiting NDR. Green line illustrates a load line represented by $R_s$. (b) Schematic of a relaxation oscillator with the NDR device marked $M$, biased with $V_{DC}$ to be consistent with the load line in (a) (i.e.: to fix the operating point within the NDR region). (c)-(e) Illustration of the working of a relaxation oscillator by displaying the current through $R_s$ and the voltage across $C_p$. (e) The waveform in (c) is illustrated with typical current flow in the circuit in (b) at different time steps (A-F). The labels from (b) are dropped for simplicity of presentation. The arrows in dark red indicate direction of current flow and two arrows appearing together indicate the effective sum of two currents. Within the ON duration, the current through $R_s$ consists of a sum of two currents (one through $C_p$ and another through $M$), as indicated. In the duration outside the ON duration, the current through $R_s$ consists of only one current (through $C_p$), as indicated (in an ideal device with no off state leakage). The length of the arrows also indicates the relative magnitudes of the currents.

## 2. Compact Model used for simulations

While periodic spiking/oscillatory behavior produced by relaxation oscillators constructed with NDR elements is well understood, here we aimed to qualitatively explain the dramatic changes in the periodic spiking behavior between the $VO_2$-only and CNT-$VO_2$ devices. The electronic transport was represented by a Schottky model (Equation S1), which produces NDR, and therefore enables multi-stability, a pre-requisite for oscillatory behavior. [*Advanced Functional Materials, 28, 1704175 (2018)*] Newton's law of cooling was used as the state equation for temperature dynamics as a lumped thermal model for the $VO_2$-only (Equation S2) and CNT-$VO_2$ (Equation S3) devices.

$$i_m = AT^2 e^{\left(\frac{\beta\sqrt{v_\mathrm{m}/d}-\phi}{kT}\right)} \tag{S1}$$

$$\frac{\mathrm{d}T}{\mathrm{d}t} = \frac{i_\mathrm{m} v_\mathrm{m}}{C_\mathrm{th}} - \frac{T-T_\mathrm{amb}}{C_\mathrm{th} R_\mathrm{th}} \tag{S2}$$

$$\frac{\mathrm{d}T}{\mathrm{d}t} = \frac{i_\mathrm{m} v_\mathrm{m} + v_m^2/R_\mathrm{CNT}}{C_\mathrm{th}} - \frac{T-T_\mathrm{amb}}{C_\mathrm{th} R_\mathrm{th}} \tag{S3}$$

Where $i_\mathrm{m}$ and $v_\mathrm{m}$ are the current and voltage through the memristive NDR device, respectively; $T$ is the absolute device temperature; $T_\mathrm{amb}$ is the ambient temperature; $A$ and $\beta$ are scaling constants; $d$ is the effective device length; $k$ is the Boltzmann constant; $\phi$ is an energy barrier; $C_\mathrm{th}$ is the thermal capacitance (effective thermal mass); $R_\mathrm{th}$ is the thermal resistance; $R_\mathrm{CNT}$ is the value of the resistor representing the CNT, taken to be 600 kΩ. Equation S3 represents both the electrical and thermal coupling between $R_\mathrm{CNT}$ and $VO_2$.

The effective circuit was represented by the schematic in Figure S2. The RC components of the top electrode (subscript TE), bottom electrode (subscript BE), and oscilloscope along with its cables (subscript Osc) were included (top and bottom are defined for ease of labelling on the schematic, and not for device structure). The intrinsic capacitor ($C_\mathrm{p}$) in parallel to $M$ and the resistor in series with a current source were also included. The equations representing the circuits, along with the transport equation and state equation for temperature were simultaneously and self-consistently solved using LT Spice. The parameters used are displayed in Table S1. $C_\mathrm{th}$ was the only parameter altered between the two simulations shown in main text Figure 3.

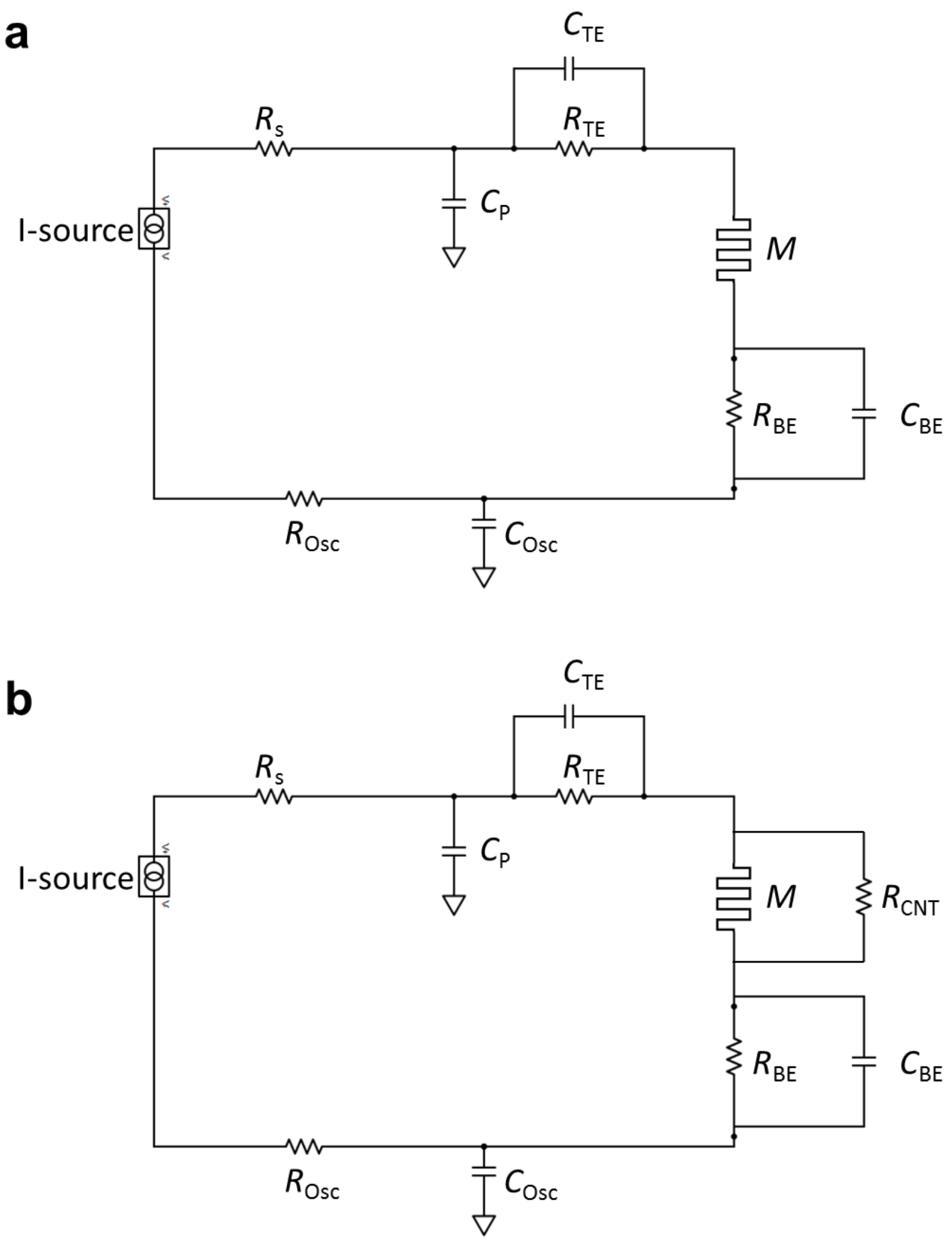

**Figure S2:** Schematic of the circuit used for simulating the periodic spiking for the (a) $VO_2$-only device, and (b) CNT-$VO_2$ device.

**Table S1:** Parameters used for simulations in main text Figure 3. Altered parameters are marked in red colored text.

| Parameter | $VO_2$-only device | CNT- $VO_2$ device |
|---|---|---|
| $T_{amb}$ | 296 K | 296 K |
| $\beta$ | $3.3 \times 10^{-4}\ \mathrm{eV \cdot m^{0.5} \cdot V^{-0.5}}$ | $3.3 \times 10^{-4}\ \mathrm{eV \cdot m^{0.5} \cdot V^{-0.5}}$ |
| $A$ | $1.7 \times 10^{-9}\ \mathrm{A \cdot K^{-2}}$ | $1.7 \times 10^{-9}\ \mathrm{A \cdot K^{-2}}$ |
| $d$ | $5 \times 10^{-6}$ m | $5 \times 10^{-6}$ m |
| $R_{th}$ | $2.5 \times 10^{8}\ \mathrm{K \cdot W^{-1}}$ | $2.5 \times 10^{8}\ \mathrm{K \cdot W^{-1}}$ |
| $\mathbf{C_{th}}$ | $\mathbf{1.2 \times 10^{-14}\ W \cdot K^{-1} \cdot s}$ | $\mathbf{5 \times 10^{-17}\ W \cdot K^{-1} \cdot s}$ |
| $C_p$ | $5.5 \times 10^{-9}$ F | $5.5 \times 10^{-9}$ F |
| $C_{TE}$ | $3 \times 10^{-11}$ F | $3 \times 10^{-11}$ F |
| $R_{TE}$ | $2.7 \times 10^{4}\ \Omega$ | $2.7 \times 10^{4}\ \Omega$ |
| $C_{BE}$ | $5 \times 10^{-16}$ F | $5 \times 10^{-16}$ F |
| $R_{BE}$ | $1 \times 10^{3}\ \Omega$ | $1 \times 10^{3}\ \Omega$ |
| $R_s$ | $3 \times 10^{5}\ \Omega$ | $3 \times 10^{5}\ \Omega$ |
| $\phi$ | 0.58 eV | 0.58 eV |
| $R_{Osc}$ | $50\ \Omega$ | $50\ \Omega$ |
| $C_{Osc}$ | $100 \times 10^{-12}$ F | $100 \times 10^{-12}$ F |
| $k$ | $8.62 \times 10^{-5}\ \mathrm{eV \cdot K^{-1}}$ | $8.62 \times 10^{-5}\ \mathrm{eV \cdot K^{-1}}$ |
| $I_{ext}$ | $9 \times 10^{-6}$ A | $9 \times 10^{-6}$ A |

We acknowledge that the model is neither rigorous nor unique, but it nonetheless illustrates the important contributions of the CNT's scaling behavior. It is possible that other combinations of the parameters might better reproduce the shape and durations of the dynamics. To demonstrate this, we provide a second set of parameters that can reasonably reproduce the shape, frequency and durations of the oscillations (Table S2, Figure S3). Here we used only one circuit (Figure S2a) for both the $VO_2$-only device and the CNT-$VO_2$ device, therefore eliminating the addition of $R_{CNT}$ upon introduction of the CNT. The following different parameters were altered to represent the introduction of a CNT: (1) $C_{th}$ was reduced upon introduction of a CNT since the CNT reduces the total volume of the material that is heated to trigger switching, and thereby the volume within which the maximum current density flows. (2) $C_p$ was reduced upon introduction of a CNT since the CNT effectively shorts most of the capacitance between the electrodes. (3) $C_{TE}$ was reduced upon introduction of a CNT because we previously discovered that the metallic CNT has a lower contact resistance to the metallic electrodes, compared to an insulating/semiconducting $VO_2$ film contacting the metallic electrodes.

We discovered that the initial spike seen in the inset of main text Figure 3c originates from the top electrode parasitic RC components (subscript TE in Figure S2). $C_p$ alters both the frequency of periodic spiking and the width of the slower transient (distinctly seen in main text Figure 3c). $C_{th}$ and $R_{th}$ also alter the width of the transients and frequency of the periodic spiking.

**Table S2:** Parameters used for simulations. Altered parameters are marked in red colored text.

| Parameter | $VO_2$-only device | CNT- $VO_2$ device |
|---|---|---|
| $T_{amb}$ | $296\ \mathrm{K}$ | $296\ \mathrm{K}$ |
| $\beta$ | $8.17 \times 10^{-5}\ \mathrm{eV \cdot m^{0.5} \cdot V^{-0.5}}$ | $8.17 \times 10^{-5}\ \mathrm{eV \cdot m^{0.5} \cdot V^{-0.5}}$ |
| $A$ | $1.7 \times 10^{-9}\ \mathrm{A.K^{-2}}$ | $1.7 \times 10^{-9}\ \mathrm{A.K^{-2}}$ |
| $d$ | $3 \times 10^{-7}\ \mathrm{m}$ | $3 \times 10^{-7}\ \mathrm{m}$ |
| $R_{th}$ | $2.5 \times 10^{8}\ \mathrm{K \cdot W^{-1}}$ | $2.5 \times 10^{8}\ \mathrm{K \cdot W^{-1}}$ |
| $\boldsymbol{C_{th}}$ | $\mathbf{1.2 \times 10^{-14}\ W \cdot K^{-1} \cdot s}$ | $\mathbf{1.2 \times 10^{-17}\ W \cdot K^{-1} \cdot s}$ |
| $\boldsymbol{C_{p}}$ | $\mathbf{5.5 \times 10^{-9}\ F}$ | $\mathbf{3 \times 10^{-12}\ F}$ |
| $\boldsymbol{C_{TE}}$ | $\mathbf{1 \times 10^{-11}\ F}$ | $\mathbf{2 \times 10^{-14}\ F}$ |
| $\boldsymbol{R_{TE}}$ | $\mathbf{2.7 \times 10^{4}\ \Omega}$ | $\mathbf{1 \times 10^{4}\ \Omega}$ |
| $C_{BE}$ | $4 \times 10^{-15}\ \mathrm{F}$ | $4 \times 10^{-15}\ \mathrm{F}$ |
| $R_{BE}$ | $1 \times 10^{3}\ \Omega$ | $1 \times 10^{3}\ \Omega$ |
| $R_{s}$ | $3 \times 10^{5}\ \Omega$ | $3 \times 10^{5}\ \Omega$ |
| $\phi$ | $0.58\ \mathrm{eV}$ | $0.58\ \mathrm{eV}$ |
| $R_{Osc}$ | $50\ \Omega$ | $50\ \Omega$ |
| $C_{Osc}$ | $1 \times 10^{-10}\ \mathrm{F}$ | $1 \times 10^{-10}\ \mathrm{F}$ |
| $k$ | $8.62 \times 10^{-5}\ \mathrm{eV \cdot K^{-1}}$ | $8.62 \times 10^{-5}\ \mathrm{eV \cdot K^{-1}}$ |
| $I_{ext}$ | $9 \times 10^{-6}\ \mathrm{A}$ | $9 \times 10^{-6}\ \mathrm{A}$ |

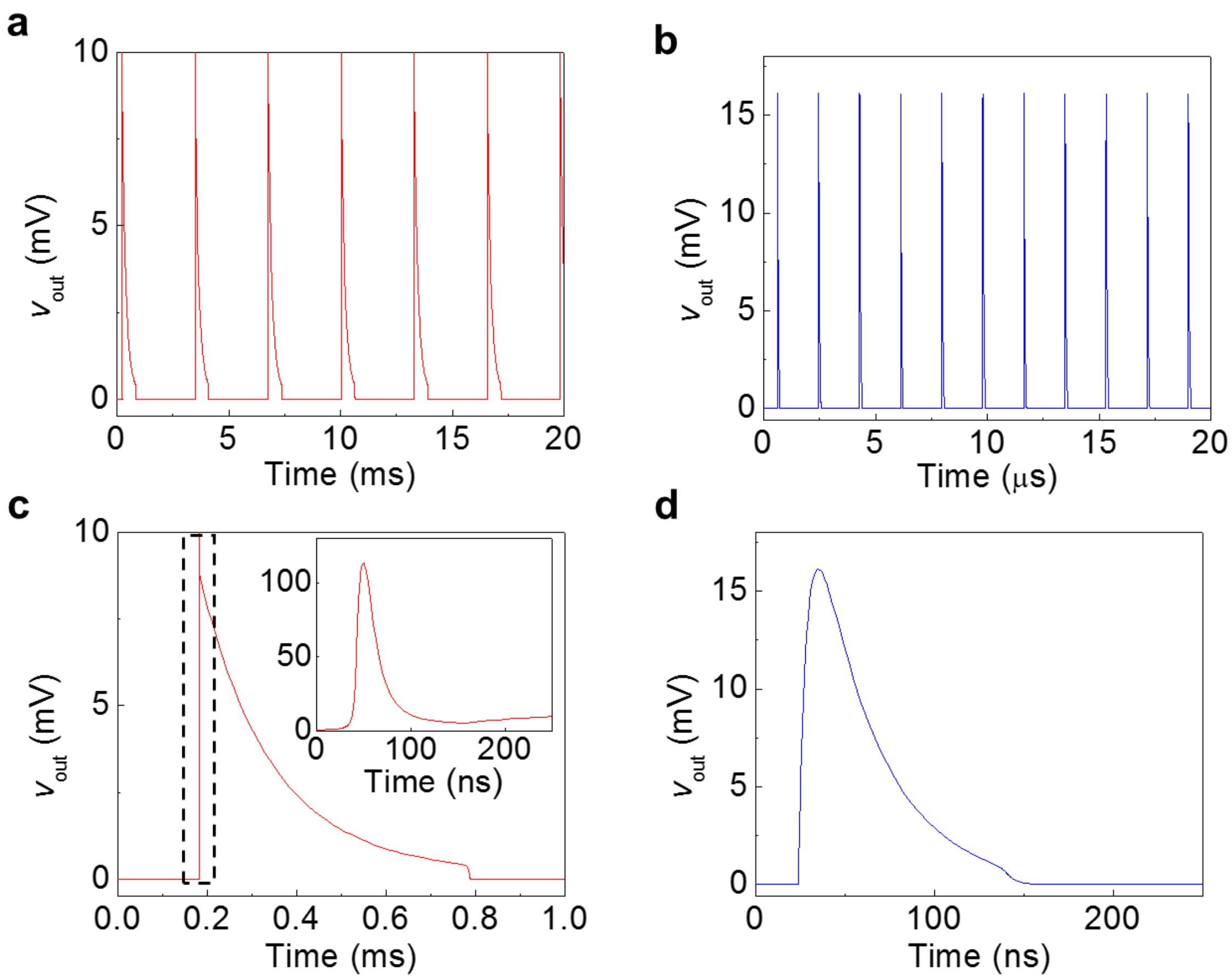

**Figure S3:** Data obtained by simulating the model represented by the parameter set in Table S2, in comparison with main text Figures 3e-3h. All abscissa (time) values may have been arbitrarily offset for clarity of presentation. All insets have the same ordinate units as the corresponding panel.

There could be further details of the model, which need to be explored in order to determine their importance in representation. It is likely that $R_{\text{th}}$ may be a function of current/power, since the CNT would act as a good heat source when the power is turned on. On the other hand, the CNT would be a better heat sink when the power is turned off, owing to its relatively high thermal conductivity. Further, the RC components of the bottom electrode (subscript BE in Figure S2) may be comparable to those of the top electrode and may even contribute distinct dynamics of their own. In addition, $R_{TE}$, $C_{\text{th}}$ and $R_{\text{th}}$ are very likely to be functions of temperature (especially since $VO_2$ undergoes a Mott transition), which further complicates the model.

## 3. Device Fabrication and Experimental Setup

$VO_2$ thin films were grown on 1 cm$^2$ single crystal $TiO_2$ (101) pieces using pulsed laser deposition (PLD). [*Science*, *339*, *1402 (2013)*] The films are smooth, with a very thin oxidation of the surface to $V_2O_5$. The $VO_2$ exhibits a $10^3$ change in resistance at transition temperatures of $T_{IMT}$ ~ 328 K.
Separately, carbon nanotubes (CNTs) were grown via chemical vapor deposition on ST-cut quartz using an ethanol source, which resulted in aligned horizontal CNTs with an average diameter of 1.2 nm. The CNTs grown were a random mixture of semiconducting and metallic CNTs, but only devices with a metallic CNT were examined in this work. Since CNTs can only grow aligned on particular substrates, they were transferred to the $VO_2$ using a gold and thermal release tape based method. [*IEEE Trans., 8, 498 (2009)*]

The $VO_2$ and CNTs are etched into stripes 2 – 10 μm wide to restrict our study to single or few CNT devices and to limit leakage current through the $VO_2$. CNTs outside the devices are removed using a light $O_2$ plasma (20 sccm, 150 mTorr, 30 W, 1 min), then the $VO_2$ is wet etched for 30 s using a 25% nitric acid solution. 50 nm thick Pd contact pads are e-beam evaporated and patterned using liftoff, with device lengths of 3 - 10 μm. Shorter devices with lengths 300 nm – 2 μm were made by adding small extensions of e-beam evaporated Pd to the previous pads using e-beam lithography.

Electrical measurements are performed in air in a micromanipulator probe station from Janis Research with a Keithley 4200-SCS parameter analyzer as a current or voltage source. All measurements are done at room temperature (~ 296 K). A resistor is added in series to the device, typically tens or hundreds of kΩ depending on the device dimensions and the presence of a CNT. Such a resistor is necessary as a current compliance that protects devices from failure in the metallic state of $VO_2$, when high power could damage a device. An ideal value is one that minimally affects the off-state characteristics or $V_{IMT}$ ($R_s$ is a small fraction of the high resistance insulating state), but that prevents permanent damage in the on state (higher than the low resistance metallic state). There is no one correct value for $R_s$, and the value of $R_s$ can be used to adjust oscillation frequency, as shown in the main text.

In order to protect devices, the resistor should be as physically close to the device as possible. When the $VO_2$ switches to high conductivity, any capacitors connected to the biased electrode will discharge (a current overshoot). This can include the capacitors intrinsic to the device and contacts (shown in the model) but can also include parasitic capacitances from the source, the cables, the probe station, etc. When these add up, the discharge can be enough to overheat and damage the device. However, if the series resistor is physically close to the device, then only the intrinsic capacitors see the change in conductance (the resistor shields the device from the other external capacitances). We placed our resistor at the end of the probe shown in Figure S4.

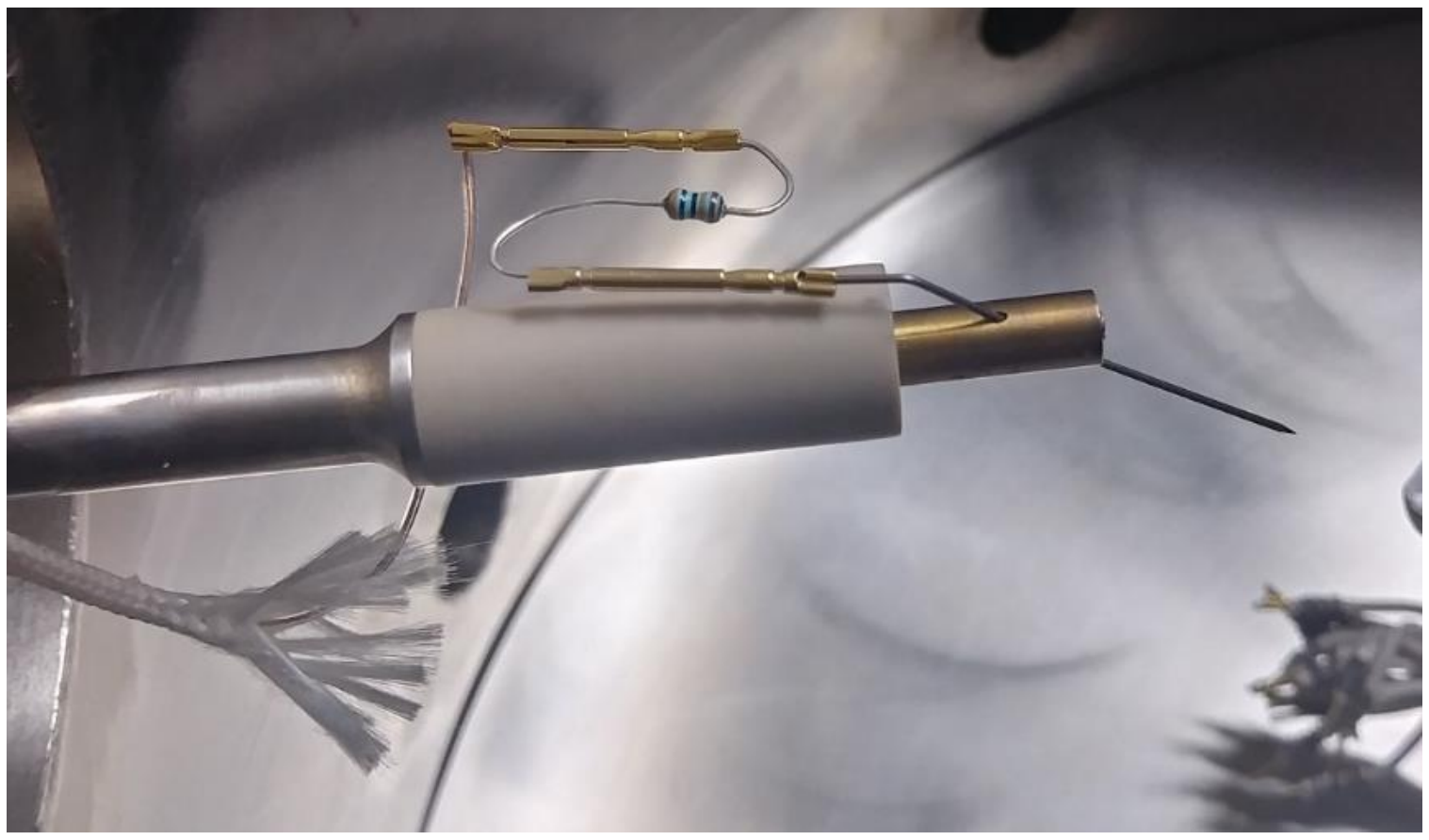

**Figure S4:** Use of a series resistor as a current compliance.

To measure oscillations, an Agilent InfiniiVision MSO710A oscilloscope or a Tektronix TDS 3014B oscilloscope is used with 50 Ω input impedance. The oscilloscope is in series with the device, at the bottom electrode side, as shown in Figure S2. Oscillations are measured while a constant voltage or current is applied from the Keithley 4200-SCS.

## 4. Additional characterization of oscillations

Sample waveforms showing repeated oscillations are shown in Figure S5, with a constant voltage applied to the circuit. Figures S5a-c are for the same CNT device at different applied voltages. As the voltage increases, the oscillation frequency increases, as does the pulse width very slightly. Figure S5d shows oscillations for our shortest CNT device, 0.3 μm long, corresponding to the shortest pulse widths and highest frequency of oscillation.

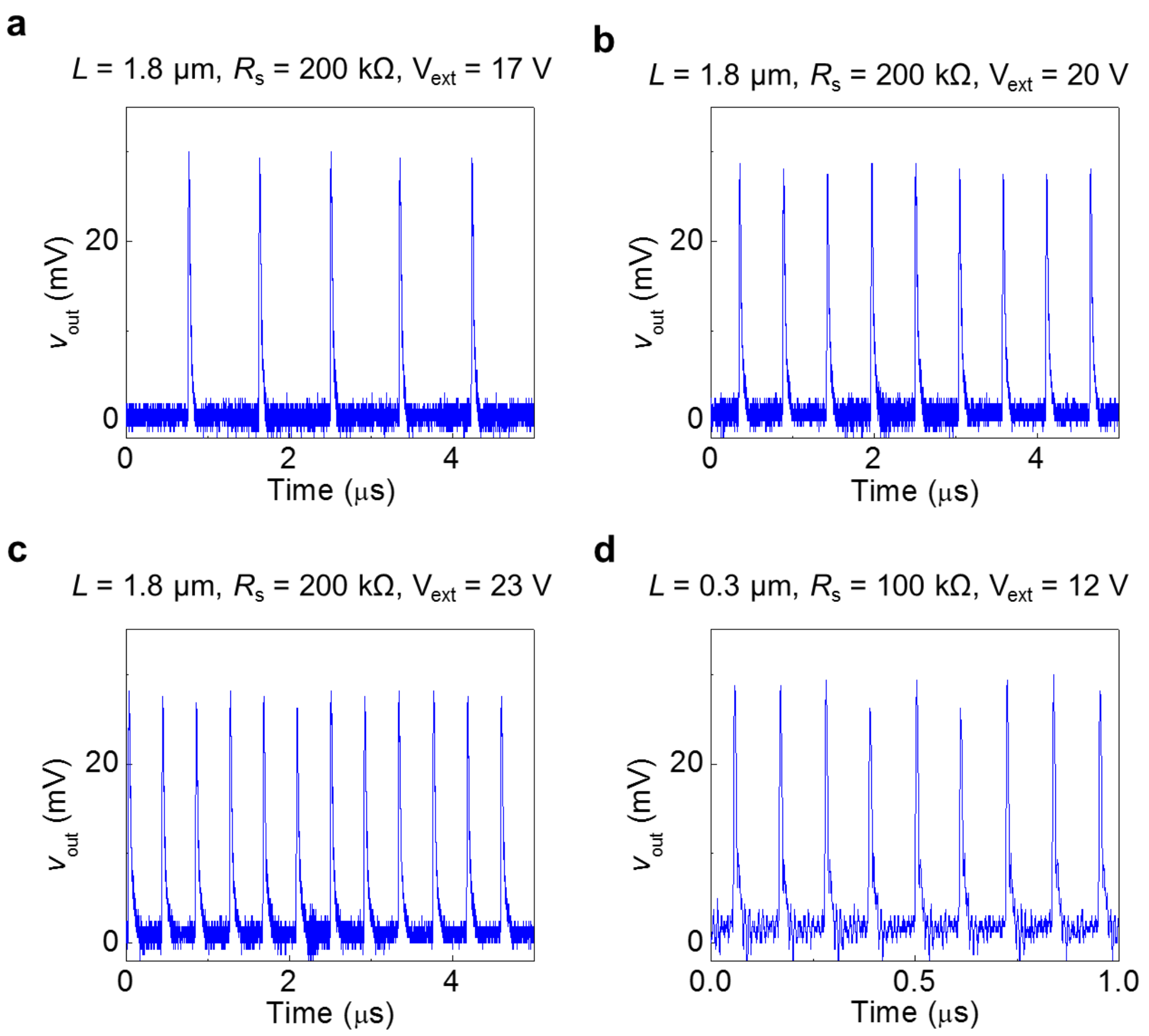

**Figure S5:** (a) – (b) Additional oscillation waveforms for a 1.8 μm long $VO_2$-CNT device with different applied voltages. (d) An oscillation waveform for a short 0.3 μm long $VO_2$-CNT device, with a 9 MHz oscillation frequency. Note the different time scale in (d).

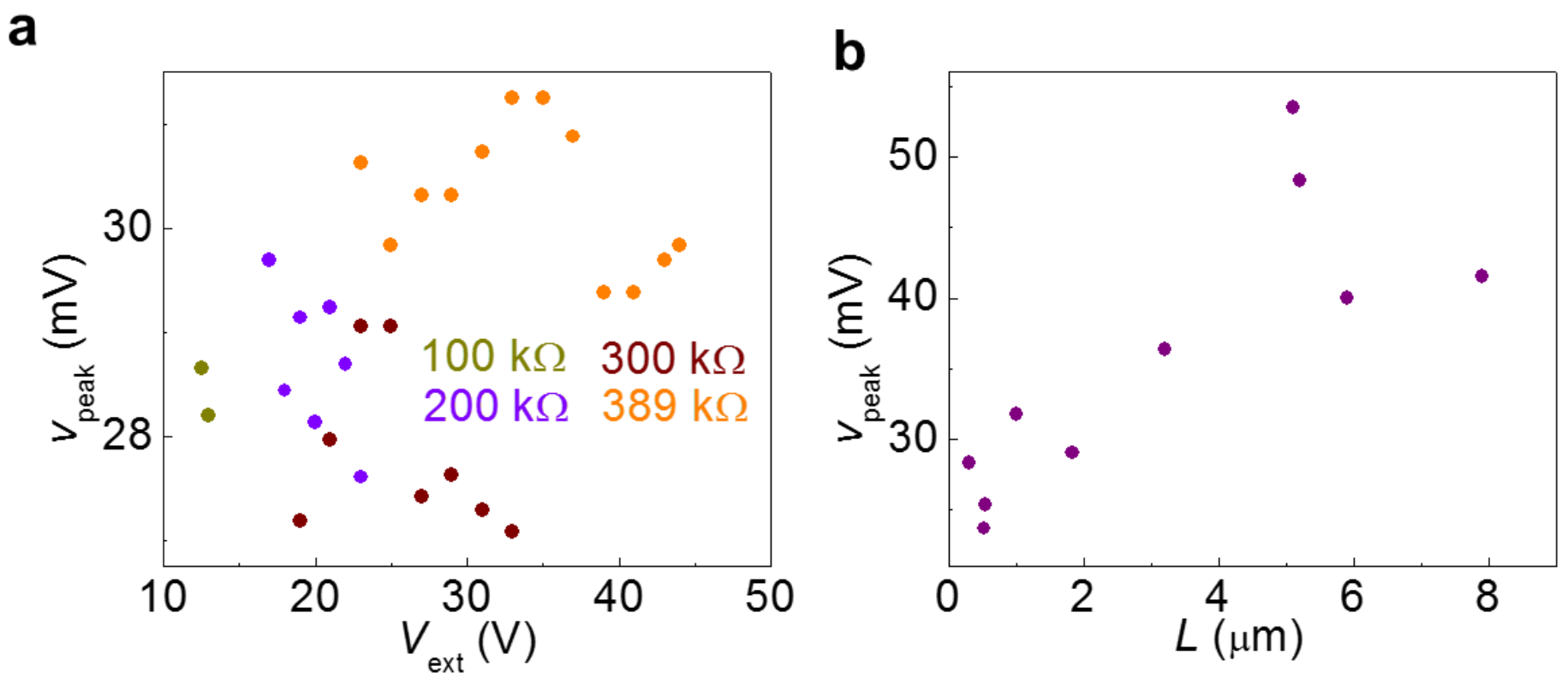

**Figure S6:** (a) Maximum pulse voltage, averaged over several pulses, for a single 1.8 μm long $VO_2$-CNT device as a function of series resistor and voltage. (b) Maximum pulse height against different lengths of $VO_2$-CNT devices.

Regardless of series resistor and applied voltage used, the height of the pulses remains the same within error (Figure S6a). The oscilloscope resolves voltages down to a ~0.6 mV discretization, but there is a few mV of noise. Any shifts in the off-state resistance or static switching voltage of the device could also lead to minor changes in pulse height.

As indicated in Figure S6b, pulse height does reduce slightly as device length decreases. This could be a result of reduced device resistance with scaling (with a lower static switching voltage and similar threshold current), resulting in lower electrical RC delays and faster pulses. It could also be the result of a reduction in the thermal capacitance of the device, since a smaller length (volume) needs to be heated by the CNT.

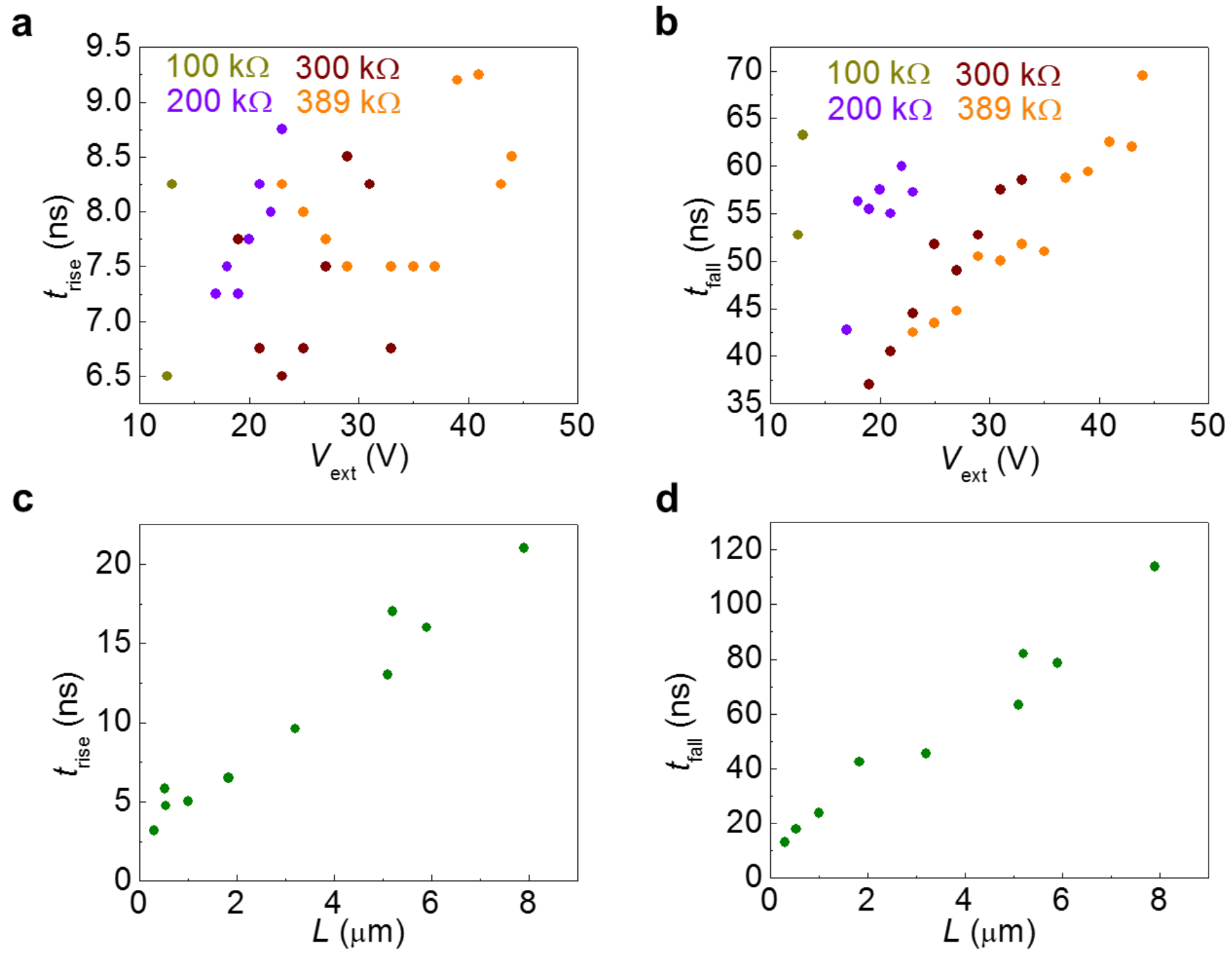

**Figure S7:** Characterization of the rise and fall times of pulses as a function of series resistance and voltage for a single 1.8µm long CNT device (a-b), and across multiple CNT devices of several lengths (c-d).

Figure S7 characterizes the rise and fall times of each pulse. Rise time is defined as the time it takes to rise from 10% to 90% of the maximum pulse voltage. Similarly, fall time is the time to fall from 90% to 10% of the maximum pulse voltage. The rise time is independent of the applied voltage or series resistor used (Figure S7a). However, the fall time increases as the applied voltage increases for a given series resistor. As a result, the total pulse width follows similar trends, increasing with applied voltage. Operating at a higher bias point means that more energy is input to the device, and it will take longer for the $VO_2$ to cool and return to its insulating state once the capacitors discharge.

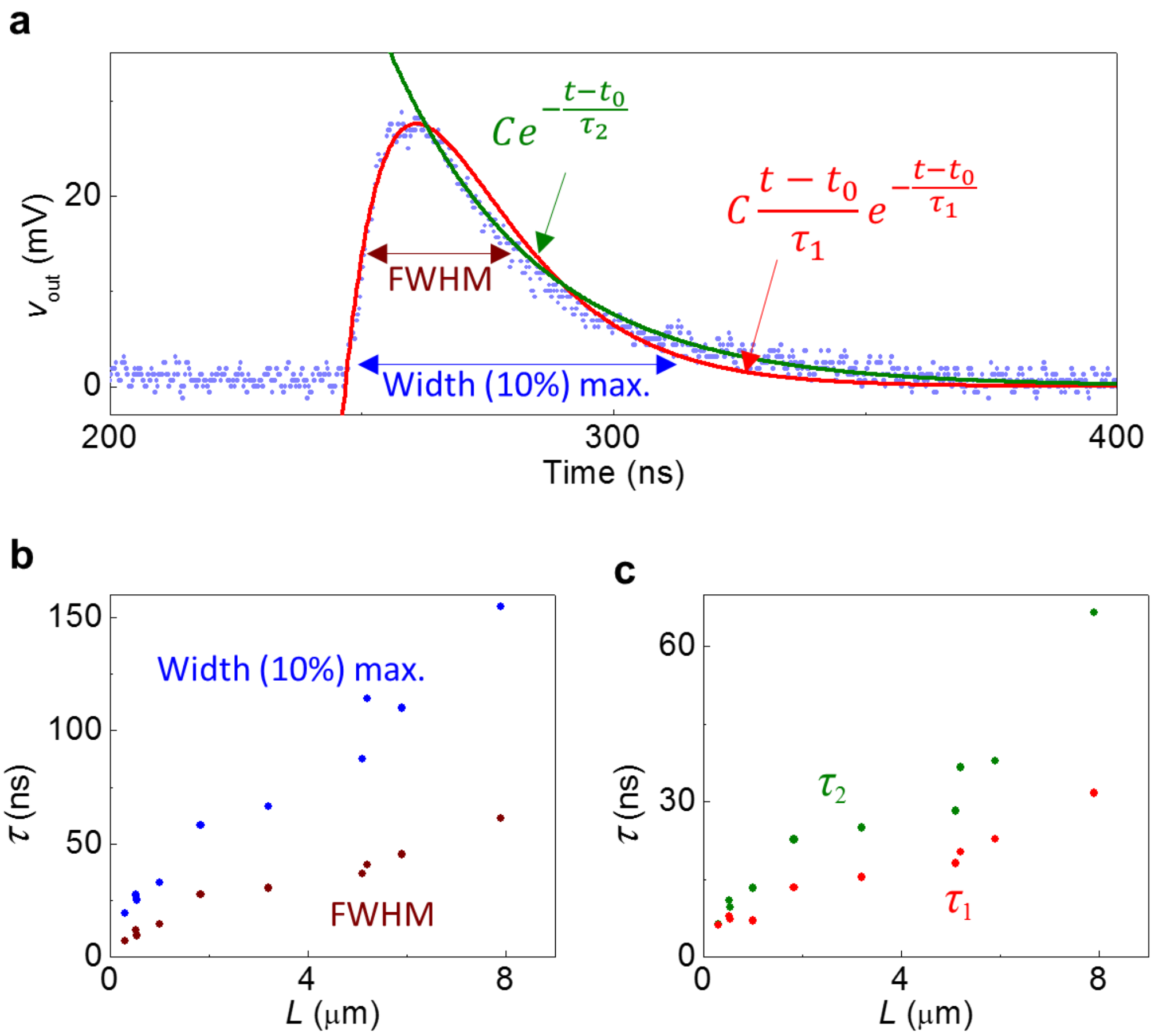

**Figure S8:** (a) Different definitions of pulse width, including the time constants of two fits. The experimental waveform shown (blue scatter) is for a 1.8 µm long CNT device with $R_s$ = 200 kΩ and an applied voltage of 20 V. (b) – (c) Scaling trends of different definitions of pulse width. The pulse width for a given device is taken as the minimum measured, at the lowest applied voltage and a large enough $R_s$ (see main text Figure 4b and S7b).

There are several ways to characterize the width of a pulse, some of which are shown in Figure S8a. The full-width half maximum (FWHM) is shown in Figure S8b versus device length, as well as a similar measure of width using ten percent of the maximum. Fitting an equation of the form $t/\tau_1 \cdot \exp(-t/\tau_1)$ to the pulse yields a time constant $\tau_1$, and fitting an exponential decay of the form exp($-t/\tau_2$) to the pulse falling edge yields a time constant $\tau_2$. A time offset $t_0$ and scaling factor C were also included. Both time constants scale similarly with length, shown in Figures S8c-d.

Linearly fitting the pulse width versus length yields a non-zero y-intercept on the order of 5 – 20 ns depending on the definition of width. This may represent a limitation of the system to measure narrow pulse widths, such as from parasitic capacitances that smear out an intrinsic sharp pulse.

**5. Using PMMA to Infer Temperature Dynamics**

To capture an qualitative imprint of the region and the dynamics of the switching of devices, we made use of a 50 nm thick poly(methyl methacrylate) (PMMA) layer capping the devices. PMMA has a low boiling point of ~523 K with an even lower reflow temperature, and has been shown to evaporate off nanowires as they are heated, with the volume of the affected PMMA increasing with local temperature. [*Nano Research 9, 2950 (2016); Nano Letters 11, 4736 (2011)*] When the device current spikes at the IMT so does the temperature, and if the time constant is long enough then the heat can be enough to cause PMMA reflow near the switching regions, preserved as a change in PMMA height after electrical switching. In other words, the dynamics of the device operation are imprinted on the PMMA. Examples of this are shown in atomic force micrographs (AFM) images Figures S9a-b for devices without and with a CNT, respectively. To observe large PMMA bumps, extra capacitance was added by connecting $R_S$ to the measurement setup with cables instead of directly attaching it to the probe the way we do for all other electrical measurements. The bumps formed from the PMMA reflow in both devices consisted of a tall spike in the center of a shorter and wider raised bubble. The tall spike in the center of the bubble represents the initial IMT location in the $VO_2$, while the shorter and wider bubble is a result of dynamical positive feedback expanding the initial IMT area while the initial current/temperature spike was decaying. These two features suggest that two time constants are present in the system, consistent with the two time constants seen electrically in each current spike.

In a $VO_2$ device without a CNT (Figure S9a), the taller PMMA bubble traces the line of a filament of least resistance between the contacts, where the device is hottest, the exact shape of which depends on the local resistivity of the $VO_2$. There is a much shorter bubble extending across the width of the device except for the very edges. This shows that at one point during the dynamics, most of the device was metallic, although once the transient was over, the metallic volume shrunk to only the central part of the device.

In a $VO_2$ device with a CNT (Figure S9b), a tall, narrow PMMA spike forms directly over the CNT, where it is hottest and switching begins. The metallic region spreads outwards creating a shorter bubble around the spike, extending ~800 nm on either side of the CNT. The bubble is much shorter and narrower than the $VO_2$ device without a CNT, as a result of much more localized, lower power transition dynamics. Although the CNT is in the center of the device in Figure S9b, the CNT and its corresponding PMMA bubble could be located anywhere on the $VO_2$ channel with the same results, including along the device edge. Using a metallic CNT to trigger IMT thus provides a mechanism of also controlling the switching location in a large device. In addition, the profiles of the PMMA distortions or bubbles qualitatively agree with the experimental and simulated results.

The lateral scale of the PMMA bubbles is affected by the parasitic capacitance, as well as $V_{IMT}$ and $R_S$. If we use the same setup as our electrical measurements with a lower parasitic capacitance ($R_S$ directly on probe, Figure S4), then no PMMA bubbles are seen in devices with a CNT, and those seen in devices without a CNT are much smaller, as shown in Figure S10a. This means the device temperature is much lower, and the switching volumes may be smaller as a result.

In devices with multiple metallic CNTs a PMMA bubble is formed around only one metallic CNT. This is shown in Figure S10b with a device that has 4 CNTs, at least two of which are metallic based on the currents measured. This indicates that the most conductive CNT triggers the IMT, unaffected by the presence of other CNTs.

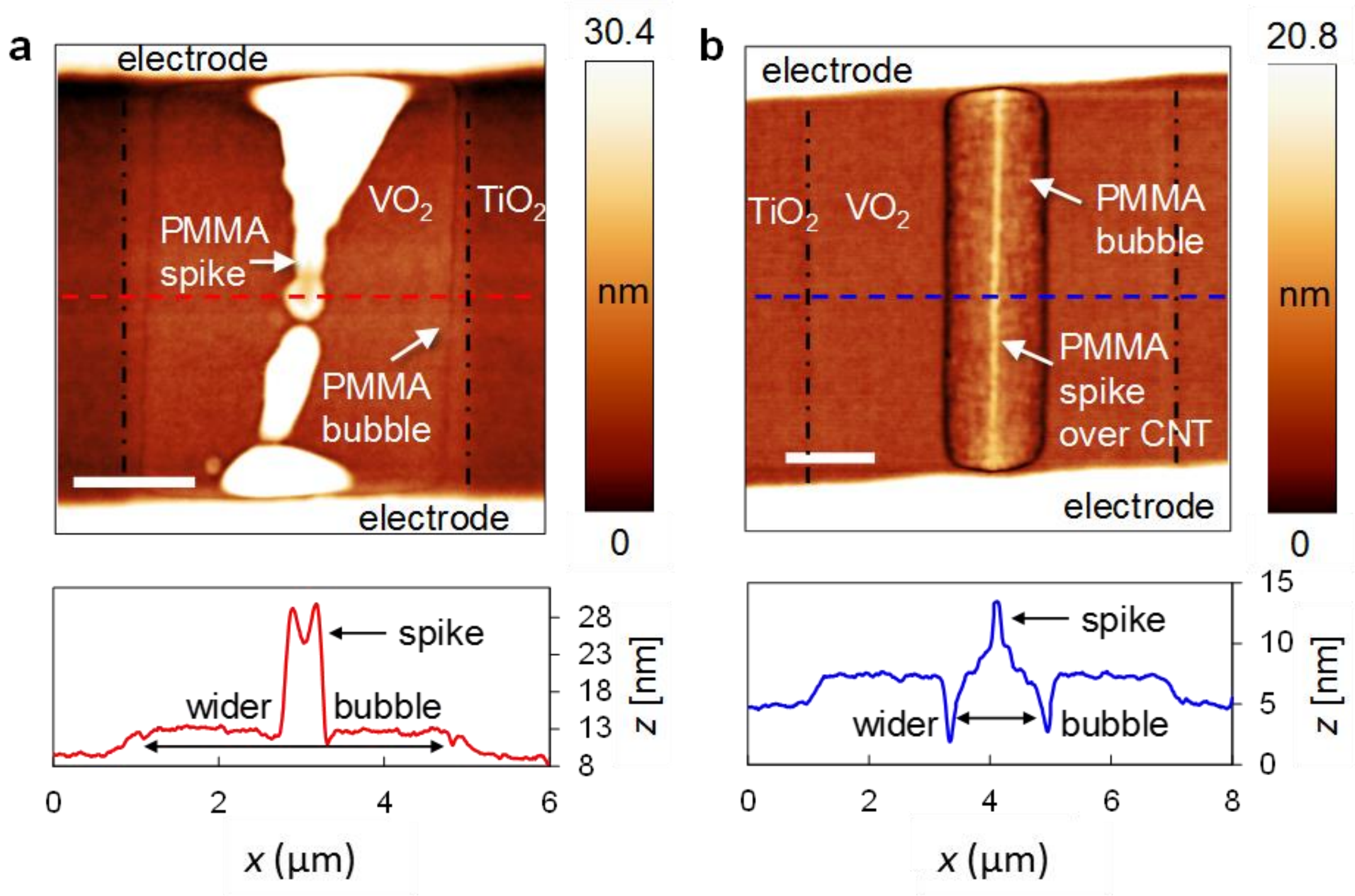

**Figure S9.** AFM images of PMMA reflow upon switching of PMMA-coated devices without (a) and with a CNT (b), due to transient spikes in temperature (current overshoot). Cabling between the device and $R_S$ is added to increase the current overshoot and visibility of reflow (by increasing the parasitic capacitances). The edges of the $VO_2$ channel underneath the PMMA are shown with dot-dash lines. Line scans of height are shown along the colored dashed lines. Scale bars are 1.5 μm.

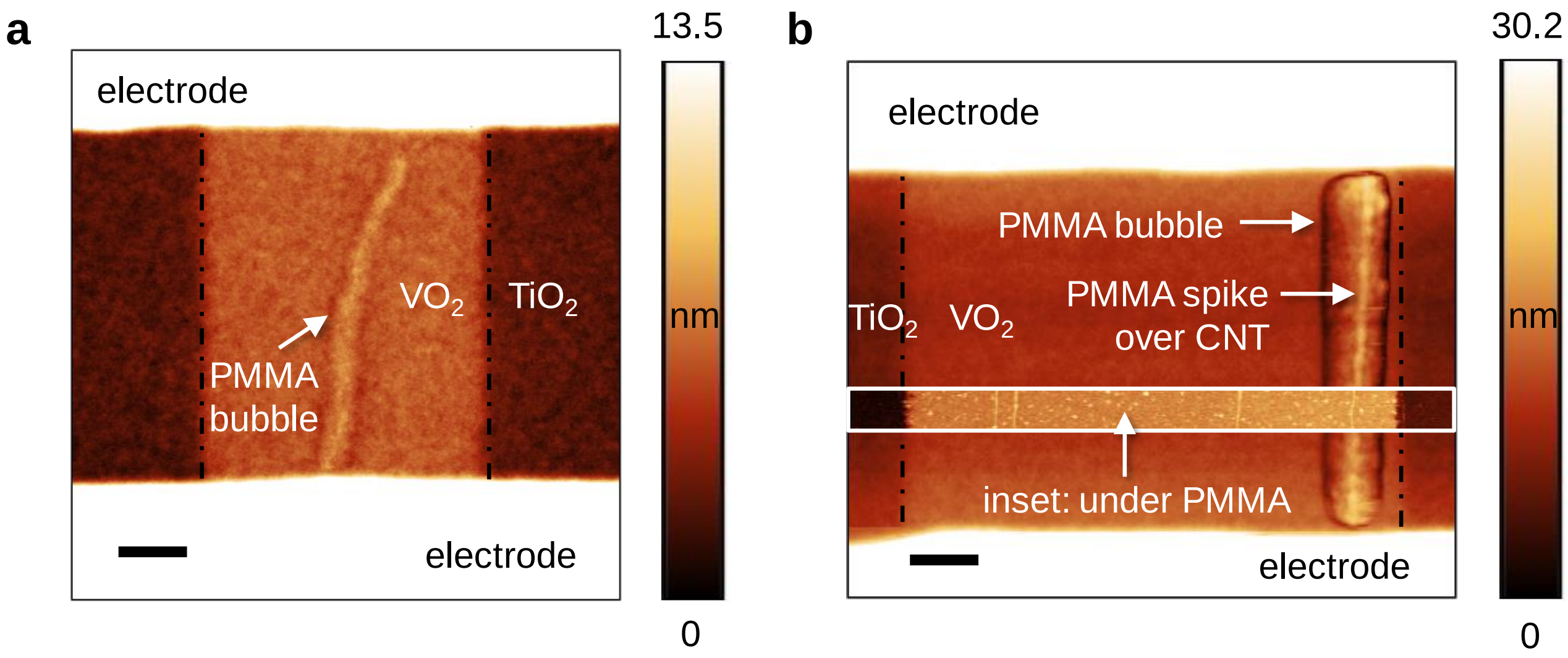

**Figure S10.** (a) AFM image of PMMA reflow upon switching of a PMMA-coated device without a CNT in the setup used for all electrical measurements (with $R_S$ as shown in Figure S4). Devices with a CNT show no visible PMMA reflow in this setup. (b) AFM image of PMMA reflow upon switching of a PMMA-coated device with multiple metallic CNTs, using the setup of Figure S9 (with extra capacitance). The CNTs cannot be seen underneath the PMMA, but the inset shows the locations of the four CNTs in the device before being coated with PMMA, at least two of which are metallic. The PMMA reflow shows switching occurs only near the most metallic CNT, and can occur near the edge of the device (marked with dot-dash lines). Scale bars are 1.5 μm.

## 6. Three-Dimensional Finite Element Simulations

This section details the simulations used to produce the results in Figure 2.

We developed a three-dimensional (3D) finite element model of our devices using COMSOL Multiphysics, which self-consistently solves electrical and thermal effects. An electrical model was used to calculate the electric field and current density distributions in the device, coupled to a thermal model that determines the temperature distributions. These are coupled via Joule heating, as well as the temperature-dependent resistivities of the CNT and $VO_2$.

Equation S4 was used to obtain the electric field and current density distributions in the device while Fourier's law of heat conduction in steady state (Equation S5) was used to obtain the local temperatures. $\kappa$ is the thermal conductivity, and σ is the temperature dependent electrical conductivity (and therefore spatially dependent). These equations are coupled via $J \cdot E$ as a Joule heating source term, with $J$ being the local current density and $E$ the local electric field.

$$\nabla \cdot (\sigma(x,y,z,T)\nabla V) = 0 \qquad \text{(S4)}$$

$$\nabla \cdot (\kappa \nabla T) + J \cdot E = 0 \qquad \text{(S5)}$$

For the $VO_2$, $\sigma(T)$ was based on measurements of $R(T)$ as a function of stage temperature. The heating branch of the $R(T)$ data was used as a look-up table, using nearest-neighbor interpolation.

The conductivity of the CNT was based on a previous model [*J. Appl. Phys. 101, 093710 (2007)*] given by:

$$\sigma_{\text{CNT}}(T, V, L) = \frac{4q^2}{h}\frac{\lambda_{\text{eff}}}{A} \quad \text{(S6)}$$

$$\text{where } \lambda_{\text{eff}} = \left(\lambda_{\text{AC}}^{-1} + \lambda_{\text{OP,ems}}^{-1} + \lambda_{\text{OP,abs}}^{-1} + \lambda_{\text{defect}}\right)^{-1} \quad \text{(S7)}$$

$\lambda_{\text{eff}}$ is an effective electron mean free path (MFP) obtained using the Matthiessen's rule. It accounts for contributions from elastic electron scattering with acoustic phonons ($\lambda_{\text{AC}}$), and inelastic electron scattering by optical phonon absorption ($\lambda_{\text{OP,abs}}$) and emission ($\lambda_{\text{OP,ems}}$). Emission is influenced by the electric field, so this term is affected by the applied voltage and CNT length, and all MFPs are a function of temperature. Values of $\lambda_{\text{OP,300}}$ = 20 nm and $\hbar w_{op}$ = 0.2 eV were used in the model.An additional scattering term for defects ($\lambda_{\text{defect}}$), with a mean free path of 0.8 µm, was included to better represent *I-V* characteristics of our imperfect CNTs. The reduction in CNT conductivity with increasing temperature results in saturation of the CNT current, described elsewhere. [*J. Appl. Phys. 101, 093710 (2007)*]

Since the CNT was centered in the device, we assumed bilateral symmetry. The CNT was approximated as a rectangular prism 1.2 nm wide and 1.2 nm tall on top of the 5 nm thick $VO_2$, spanning the entire length of the device and underneath the 50 nm thick Pd electrodes. The simulated $TiO_2$ substrate could be reduced to ~ 2 µm (unlike the ~500 µm experimental $TiO_2$) for $VO_2$-CNT devices, and still be sufficient to capture its thermal resistance, most of which occurred at the thermal constriction near the CNT. For $VO_2$-only devices, where there was significantly larger volume being heated, a 15 µm thick $TiO_2$ substrate needed to be simulated to accurately determine the device temperature. The simulated $VO_2$ devices with and without a CNT were 4 µm wide and 5 µm long. A 200 kΩ series resistor $R_S$ was modeled as a 50 nm thick resistive layer on top of one of the contacts, with a resistivity chosen to yield a total resistance of 200 kΩ. The $R_S$ layer did not thermally interact with the device. Including the extra width of $TiO_2$ beyond the $VO_2$ width made no significant difference to devices with a CNT (wherein the $VO_2$ edges are near room temperature and were far from the CNT heating), but a large width of $TiO_2$ is needed to fully capture the thermal profile in $VO_2$-only devices. Material parameters are listed in Table S3.

The top of one contact was grounded, and the top of the other (or the top of $R_S$, if applicable) was set at a constant voltage. Electrical contact resistance was simulated on internal boundaries between CNT/Pd (25 kΩ), CNT/$VO_2$ (100 kΩ) and $VO_2$/Pd. The $VO_2$/Pd contact resistivity was set to $0.8\times10^{-3}$ $\Omega\cdot\text{cm}^2$ at room temperature and scaled with temperature in the same way as the $VO_2$ resistivity (by using the normalized heating branch of experimental $R(T)$). [*J. Appl. Phys. 112, 124501, (2012)*] Other boundaries were modeled as electrically insulating.

The bottom of the $TiO_2$ was fixed at room temperature (296 K). The top of the device was assumed to be thermally insulating (adiabatic), and the sides of the device had open boundary conditions. The Pd resistivity was obtained from measurements of our films, and its thermal conductivity was estimated from the Wiedemann-Franz law. Thermal boundary resistance was simulated on all interior boundaries. These interfacial resistances cannot be ignored, since they limit the flow of heat (especially from the CNT to the $VO_2$ but also from the $VO_2$ to the $TiO_2$ substrate), impacting the local $VO_2$ temperature and therefore the switching voltage. The CNT/$VO_2$ interface was set to $5\times10^{-9}$ $\text{m}^2\text{KW}^{-1}$, the $TiO_2$/$VO_2$ interface to $8\times10^{-9}$ $\text{m}^2\text{KW}^{-1}$, and all other interfaces to $10^{-8}$ $\text{m}^2\text{KW}^{-1}$.

All simulations were solved with a segregated approach, which solves each physics (temperature and voltage) sequentially, repeating until convergence was achieved. The matrix equation containing the description of the physics (Equations S4-S5) on each mesh element was solved using a direct approach (as opposed to iterative) with the MUMPS solver. The temperature was solved as the first step, followed by the electric potential. The solution for a given bias point was used as the initial conditions for the next voltage.

**Table S3.** Material properties used in simulation

| | **σ (Sm$^{-1}$)** | **κ (Wm$^{-1}$K$^{-1}$)** |
|---|---|---|
| **$TiO_2$** | $10^{-7}$ | 8 |
| **$VO_2$** | Function of $T(x,y,z)$ spanning 80 to $2\times10^{6}$ | 5 |
| **CNT** | Equations S6, S7 | 600 |
| **Pd** | $2.9 \times 10^{6}$ | 23 |

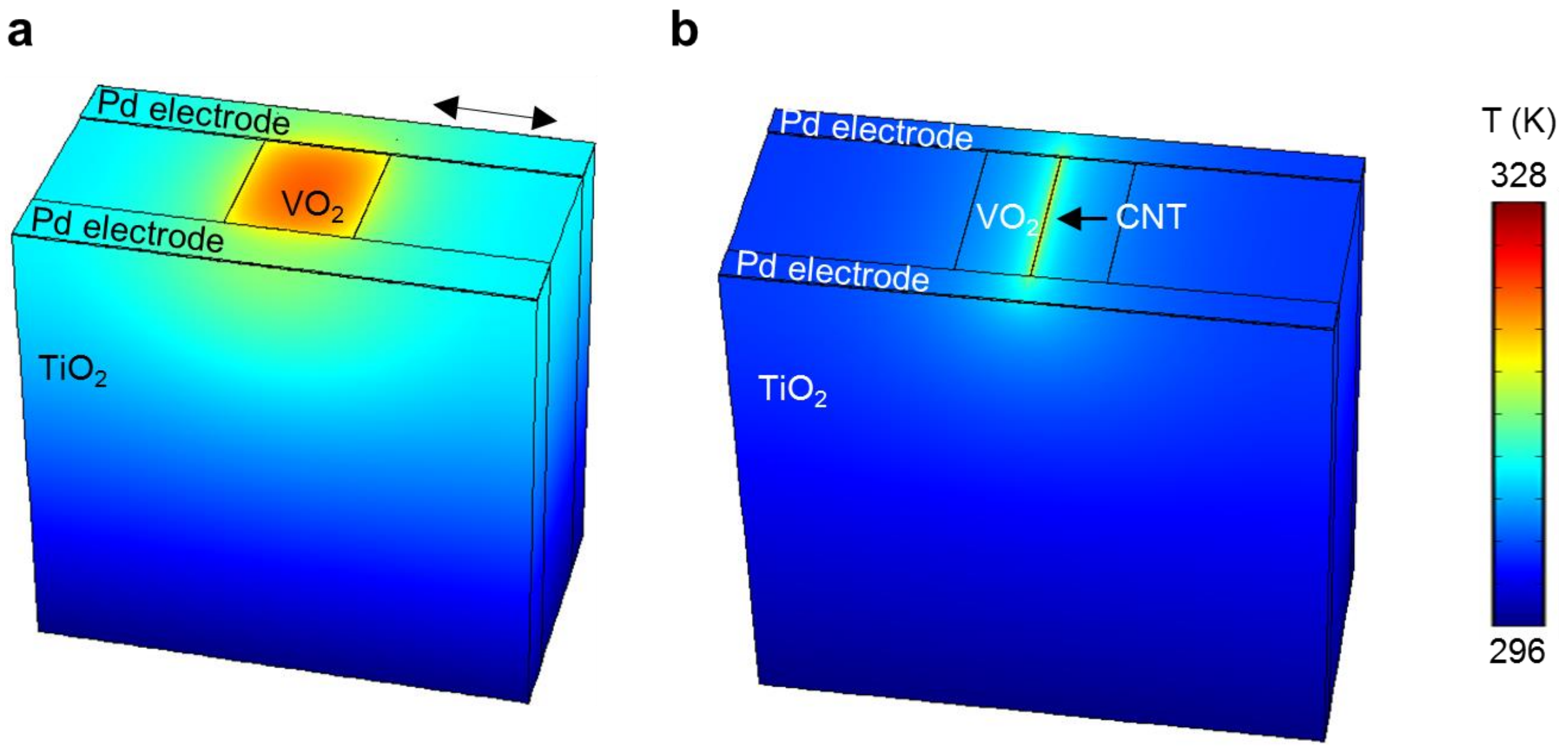

**Figure S11.** Three-dimensional views of the simulated temperature maps presented in Figure 2 for (a) the $VO_2$-only device and (b) a CNT-$VO_2$ device. Scale bar is 5 µm.